\documentstyle[aps,floats,prl,epsfig]{revtex}

\begin{document}
\twocolumn[\hsize\textwidth\columnwidth\hsize
\csname @twocolumnfalse\endcsname
\title{Pseudoenergy, Superenergy, and Energy Radiation via
       Gravitational Waves in General Relativity}
\author{Yury E. Pokrovsky}
\address{RRC "Kurchatov Institute", Moscow 123182,
Russia\footnote{E-mail: pokr@mbslab.kiae.ru}}
\date{\today}
\maketitle
\begin{abstract}
For slowly spinning matter the rate of energy loss via radiation of
gravitational waves is estimated in General Relativity (GR) within
a generally covariant superenergy approach. This estimation differs
from Einstein's Quadrupole Formula (EQF) by a suppression factor
($\Pi\ll1$). For a symmetric two-body-like distribution of scalar
matter $\Pi$ is estimated to be $\ll(v/c)^2(r/R)^2$, where $v$ is
orbital velocity of the bodies, $c$ - velocity of light, $r$ -
radius of each body, and $R$ -- the inter-body distance. This
contradiction with EQF is briefly discussed.
\end{abstract}
\vskip2pc]


Until the direct detection of the gravitational waves (GW) produced by binary
neutron stars or other astro\-physical and cosmological objects
(which is expected as highly probable in the nearest decade), the
measurements of orbital damping of the binary pulsar PSR1913+16
\cite{HulseTaylor1975,Taylor1994}, in excellent agreement with
Einstein's Quadrupole Formula
(EQF)~\cite{Einstein19161918}, are usually considered the
ultimate indirect test of the energy radiated in General Relativity
(GR)~\cite{Einstein1915} by any moving matter.

The generation of gravitational radiation in GR is a long-standing
problem that dates back to the first years following the
publications of GR \cite{Einstein19161918,Eddington1924}.
There followed a lengthy debate about whether gravitational waves
are real or an artifact of general coordinate invariance. The fact
that GW are real has been confirmed by coordinate free
theorems \cite{Bondi1957,BondiBurgMetzner1961,Sachs1962} and by
short-wave analysis \cite{Isaacson1968}.

Quantitative estimations of energy radiation via GW are based on
(or agree with) the well known Einstein's Quadrupole Formula which
is derived for pseudoenergy\footnote{In GR the gravitational field
(Levi-Civita connection coefficients $\Gamma^i_{kl}$) does not
possess any energy-momentum tensor but, as a consequence of the
Einstein Equivalence Principle (EEP), it only possesses the
so-called "energy-momentum pseudotensors" ($t^{\mu\nu}$). The
gravitational energy-momentum (and gravitational angular momentum)
pseudotensors, as being functions of $\Gamma^i_{kl}$ (the total
gravitational strengths) describe the energy-momentum of the total
gravitational field, which is a combination of the real
gravitational field (for which the Riemann curvature tensor
$R_{iklm}\ne0$) and the inertial forces field (for which
$R_{iklm}=0$). The inertial forces field is generated by the
coordinates used. This is also a consequence of the {EEP}.}
radiation rate ($p\dot E$). Then the generally covariant energy
radiation rate ($\dot E$) is estimated by the noncovariant $p\dot
E$ in the transverse-traceless (TT) gauge. The leading order of the
post-Newtonian (PN) expansion yields
\begin{equation}
p\dot E=
\frac{G}{45}
\left(\stackrel{...}{D}^{ik}\right)^2,
\label{EQF}
\end{equation}
where $G$ is Newton's gravitational constant\footnote{Throughout
this paper the units, in which the light velocity $c=1$ and reduced
Planck constant $\hbar=1$, are adopted.}, and
\begin{equation}
D_{ik}=\int T^{00}\left(3x_ix_k-\delta_{ik}r^2\right)
\label{QMom}
\end{equation}
is the time-dependent, untraced mass quadrupole tensor with mass
density $T^{00}$ given by the energy-momentum tensor of matter
$T^{\mu\nu}$ \cite{LandauLifshitz1980}.

In spite of the generally noncovariant relation $\dot E=p\dot
E$, after the discovery of binary pulsar PSR1913+16
\cite{HulseTaylor1975}, formula (\ref{EQF}) with next-to-leading
order PN contributions (see \cite{WillWiseman1996,PatiWill2000,%
Blanchet2000}, in particular) is widely accepted for GR estimations
of GW radiation, in part because of theoretical works designed to
shore up its foundations \cite{PatiWill2000,Blanchet2000,%
Ehlers1976,ChristodoulouSchmidt1979,WalkerWill1980,Anderson1980,%
IsaacsonWellingWinicour1984,Damour1984} , but mostly because of its
excellent agreement with binary pulsar data
\cite{Taylor1994,Stairs1998}.

But an empirical agreement does not imply conceptual adequacy; the
covariance in theoretical derivations motivated by many experiments
may be more important for the theoretical understanding of the
nature of gravitation than a particular agreement with data.

Gauge invariance is an important property of the wave equation and
energy-momentum tensor for gravitational waves in GR. The usual gauge
transformations used for gravitational waves are related to
coordinate transformations and gauge invariance implies invariance
under these transformations. In general the wave equation for weak
gravitational waves is approximately gauge invariant only for high
frequency waves \cite{Isaacson1968}. The effective energy-momentum
tensor is gauge invariant to leading order only if it is averaged
over a region of spacetime whose scale is large compared to the
wavelengths of the waves. In other cases, such as low frequency
gravitational waves where the averaging is done over time, either
the wave equation, the energy-momentum tensor, or both, are in
general not gauge invariant \cite{Anderson1997}.

Another alternative is to define gauge invariant effective
energy-momentum tensors for gravitational waves and other
gravitational perturbations in almost all situations of interest
\cite{Anderson1997} if only those gauge transformations are used
which change the perturbed geometry, but leave the background
geometry alone. This point of view allows considering the general
covariance for all approaches to energy radiation based on the so
called gauge invariant variables in GR \cite{Isaacson1968,%
BrillHartle1964,Burnett1988,Moncrief1974,Efroimsky1992,%
ShibataUryu2002}. Attempts to quantify the backreaction effects of
GW in GR by defining a gauge invariant effective energy-momentum
tensor for the waves were first done in \cite{BrillHartle1964} and
later in \cite{Isaacson1968} for high frequency gravitational waves
in a vacuum. Another effective energy-momentum tensor for high
frequency waves, both in a vacuum and in spacetimes containing
classical matter, was defined in \cite{Burnett1988}. An extension
of \cite{Isaacson1968} in order to include lower frequency waves
and spacetimes containing classical matter has been done in
\cite{Efroimsky1992}. It should be noted that in each case
mentioned above a procedure violating general gauge invariance
was used: either some sort of averaging procedure, or a
procedure of reduction of the gauge group that gives similar
results to an averaging procedure \cite{Anderson1997}.

In many known calculations with GW, the background spacetime
manifold is considered non dynamical and, moreover, flat. This
means a partial gauge fixing in order to work only with systems of
coordinates such that the background metric takes the Minkowski
form. What is left from the general gauge group after this gauge
fixing is just the Poincar\'e group. Thus, the reduction of the
gauge group means the transition from a general invariant theory of
gravitation to a Poincar\'e invariant one \cite{Pons2001} which is
a theory of tensor gravitational field.

This reduction of the gauge group is a possible reason why some
formally independent calculations of GW radiation in the TT gauge
(\cite{BreuerChrzanowksiHughesMisner1974,%
BreuerVishveshwara1973,BreuerRuffiniTiomnoVishveshwara1973} and
\cite{ShibataUryu2002}\footnote{In these numerical calculations the
gauge conditions are approximately transverse and traceless in the
wave zone.}) estimate the rate of energy loss in agreement with
each other and with EQF: the energy-momentum pseudotensor
transforms like the tensor under transformations of the Poincar\'e
group, as do the gauge invariant variables introduced with the
partial fixing of the general gauge group. Namely, because of the
partial fixing of the gauge group, all these estimations correspond to the
tensor field gravity in Minkowski space-time and therefore they are
close to each other. Unfortunately, all these estimations do not
correspond exactly to GR.

A classic point of view \cite{Pirani1957,Lichnerowicz1958,%
Lichnerowicz1959} is that the Riemann curvature tensor
($R_{\mu\nu\alpha\beta}$) has to play the main role in the
definition of gravitational radiation for the exact solutions to
GR: only the covariant expressions dependent on
$R_{\mu\nu\alpha\beta}$ may be used to get real information on
gravitational energy-momentum and angular momen\-tum in arbitrary
admissible coordinates.

The canonical superenergy \cite{Bel1959} and angular
super\-momentum \cite{Garecki1999} tensors are exactly this type of
quantities. They extract covariant information (hidden in the
pseudotensor ($t^{\mu\nu}$) \cite{Garecki2001}) about the real
gravitational field, and can be expressed through
$\partial_{\alpha\beta}t_{\mu\nu}\sim R_{\mu\nu\alpha\beta}^2$
\cite{MisnerThorneWheeler1973}.

Calculations of the superenergy radiation rate ($s\dot E$) can be
made by using conservation of the total superenergy introduced for
a massive scalar matter field ($\phi$) with metric gravitation
\cite{Senovilla1999}.

As a contribution to debates in \cite{WalkerWill1980},
\cite{EhlersRosenblumGoldbergHavas1976}, \cite{Thorne1980},
\cite{Loinger1999,Loinger2000}, \cite{Rosenblum1978},
\cite{Cooperstock19771980APJL1980PR}, this paper is an attempt to consider the
problem of energy radiation by scalar matter in GR within a
generally covariant super\-energy approach \cite{Senovilla1999}
with the conservation of total superenergy and a covariant relation
between the energy-momentum and superenergy-supermomentum tensors.


Let's start from an approximate two-body-like solution to the
Einstein-Klein-Gordon equation\footnote{Following
\cite{MisnerThorneWheeler1973,Thorne1980}, the convention is for Greek
indices to run over four time-space values 0, 1, 2, 3, while
Latin indices run over three spatial values 1, 2, 3; commas and
$\partial$ denote partial derivatives with respect to a chosen
coordinate system, while semicolons and $\nabla$ denote covariant
derivatives; repeated indices are summed over.}
\begin{equation}
R_{\mu\nu}=8\pi G
\left(\phi_{;\mu}\phi_{;\nu} +
\frac{1}{2}m^2\phi^2g_{\mu\nu}\right),
\label{EKG}
\end{equation}
where $m$ is the scalar field mass.


For semi-quantitative estimations let's approximate solutions to
(\ref{EKG}) by an extended, two-body-like distribution of the
scalar matter field $\phi(t,x,y,z)$ chosen in a form of fuzzy
surface ellipsoid (with major semi-axis $A$, $B$, $C$) which is
slowly spinning at angular velocity $\omega$. In order to study the
sensitivity of thus obtained results to deviation of $\phi$ from
the exact solution to (\ref{EKG}), let's consider two different
coordinate depen\-dences for $\phi(t,x,y,z)$ chosen in exponential
$\phi_{E}(t,x,y,z)$ or Gauss $\phi_{G}(t,x,y,z)$ forms\footnote{
In the $x-y$ rotation plane
$\phi_{E}\sim\phi_{1}$, $\phi_{G}\sim\phi_{2}$, where
\begin{eqnarray}
\phi_{n}=
\exp\Biggl(\Biggr.&-&
\left(\frac{|x\cos(\omega t)+y\sin(\omega t)|}{A}\right)^n
\nonumber\\ &-&
\left(\frac{|x\sin(\omega t)-y\cos(\omega t)|}{B}\right)^n
-\left(\frac{|z|}{C}\right)^n
\Biggl.\Biggr).
\label{SFE}
\end{eqnarray}}.
Neglecting gravitational mass shift, let's normalize both
distributions to the rest mass of the matter ($M$) at $\omega=0$.
For thus defined $\phi_{E}$ and $\phi_{G}$ the energy and
superenergy densities of the matter at $A\gg B\sim C$ are
two-body-like (Fig.1). Although, $\phi_{E}$ and $\phi_{G}$ neglect
radial contractions of each body under the gravitational attraction,
any radial movement of matter does not radiate the tensor metric
waves and is beyond the subject of this paper.

\begin{figure} [ht]
\vspace{4cm}
\includegraphics{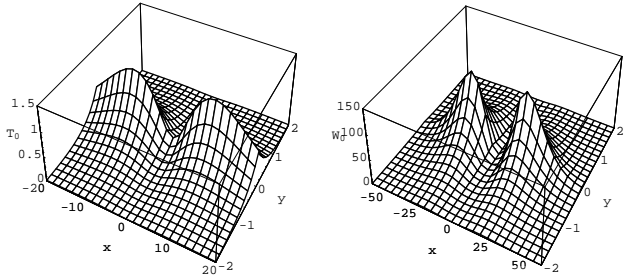}
\caption{Two-body-like energy $T_0$ (left) and super\-energy $W_0$
(right) densities projected on the rotation plane for $\phi_E$ at
$\omega=0$, and $A:B:C=10:1:1$.}
\label{Fig1}
\end{figure}


In the TT gauge small deviations ($h^{\alpha\beta}$) of metric tensor
$g_{\mu\nu}(x)$ from its Minkowski form ($\eta^{\alpha\beta}$)
yield \cite{LandauLifshitz1980}
\begin{equation}
h_{ik}=-
\frac{2G}{3R}
\left(\stackrel{..}{D}_{ik}\right)_{ret}.
\label{QMW}
\end{equation}

Instead of the non\-covariant calculations with the energy-momentum
pseudo\-tensor and gauge invari\-ant variables, let's calculate the
evolution of the scalar matter ($\cal S$) \cite{Senovilla1999} and
metric gravitation ($B$) \cite{Bel1959} terms of the
conserving\footnote{A triple contraction of
$W_{\alpha\beta\lambda\mu}$ with any three Killing vectors
$\xi^{\beta}_1\xi^{\lambda}_2\xi^{\mu}_3$ (or three copies of the
same if only one Killing vector exists) satisfy the differential
conservation law \cite{Senovilla1999}
\begin{equation}
\left[W_{\alpha\beta\lambda\mu}
\xi^{\beta}_1\xi^{\lambda}_2\xi^{\mu}_3\right]^{;\alpha}=0.
\label{WaC}
\end{equation}}
covariant superenergy tensor\footnote{Because the exact
conservation of $W$ (\ref{WaC}) takes place only for the metric
tensor and matter field satisfying (\ref{EKG}), all calculations
should not be too sensitive to a reasonable variation of the
approximate solutions mentioned above.}
($W_{\alpha\beta\lambda\mu}$) \cite{Senovilla1999}
\begin{equation}
W_{\alpha\beta\lambda\mu} = {\cal
S}_{(\alpha\beta\lambda\mu)}+B_{(\alpha\beta\lambda\mu)}.
\label{Wablm}
\end{equation}

In weak field approximation the calculations can be significantly
simplified \footnote{For the goal of this paper the most reasonable
choice of Killing vectors is the 4-velocity $u$ of an observer at
rest relative to the radiating matter: $u=(1,0,0,0)$.}, because the
generally covariant superenergy density ($W$) and flux ($P_i$) can
be chosen as $W=W_{0000}$, and $P_i=W_{i000}$.


In order to compare rates of the superenergy and pseudoenergy
losses, let's substitute $\phi_{E}$ or $\phi_{G}$ into (\ref{QMW}).
This yields simple analytical expressions for corresponding fluxes.
At large distances from the source of GW these fluxes are well
known as the Bel-Robinson supermomentum ($P\to
P^{B-R}_{E,G}$)\cite{Bel1959} and Landau-Lifshitz pseudomomentum
($P^{L-L}_{E,G}$) \cite{LandauLifshitz1980}.

In the two-body-like case ($A\gg B=C$) calculations of the
z-components of the fluxes generated by matter spinning in the x-y
plane yield
\begin{equation}
P^{B-R}_E=4\frac{G M^2}{\pi R^2}A^4\omega^8,~~~
P^{B-R}_G=2\frac{G M^2}{\pi R^2}A^4\omega^8,
\label{BelFlux}
\end{equation}
\begin{equation}
P^{L-L}_E=\frac{G M^2}{\pi R^2}A^4\omega^6,~~~
P^{L-L}_G=\frac{G M^2}{2\pi R^2}A^4\omega^6.
\label{PEF}
\end{equation}
Expression (\ref{BelFlux}) differs from (\ref{PEF}) by the factor
$4\omega^2$, and the superenergy conservation law (\ref{WaC})
results in a covariant quadrupole formula which differs from
(\ref{EQF}) by $4\omega^2$:
\begin{equation}
\dot W\sim4\omega^2\frac{G}{45}
\left(\stackrel{...}{D}^{ik}\right)^2.
\label{WQF}
\end{equation}


Finally, $E$ and $\dot E$ can be
covariantly related to $W$ and $\dot W$.
Substitution of $\phi_{E}$ or $\phi_{G}$ into $T^{00}$ and
$W^{0000}$ relates the total energy ($E_{E,G}$) and
superenergy ($W_{E,G}$). Neglecting the small high order
terms ($\omega^4 A^4 \sim v^4\ll\omega^2 A^2\sim
v^2$) in the two-body-like case ($A\gg B=C$) yields the
following covariant relations:
\begin{equation}
E_E=\frac{1}{2m^2}W_E+\frac{1}{2}M,~
E_G=\frac{1}{3m^2}W_G+\frac{2}{3}M.
\label{WEE}
\end{equation}
The reasonably small difference between $E_E$ and $E_G$ roughly
approximates possible deviations from the exact result. Because
these deviations are not signifi\-cant for semi-quantitative
estimations, only the exponential case is considered below:
\begin{equation} {\dot E}=\frac{1}{2m^2}{\dot W}.
\label{dotEdotW}
\end{equation}


Expressions (\ref{BelFlux}), (\ref{PEF}), and
(\ref{dotEdotW}) together yield the generally covariant quadrupole
formula
\begin{equation}
\dot E=
\frac{G}{45}\Pi
\left(\stackrel{...}{D}^{ik}\right)^2,
\label{GIQF}
\end{equation}
which differs from (\ref{EQF}) by a suppression factor
\begin{equation}
\Pi\approx2\frac{\omega^2}{m^2},
\label{SF}
\end{equation}
where $m$ (the scalar field mass) is yet an arbitrary parameter.

To be consistent with the classical solutions to the problem of GW
radiation, $m$ should be large enough: the scalar particle Compton
wavelength ($1/m$) should be $\ll A, B, C$, and correspondingly $m\gg m_0$, where
the energy of quantum fluctuations $m_0$ is estimated by the
inverse values of $A$, $B$, $C$. Therefore, in accordance with
(\ref{SF}) the upper limit of the energy loss rate can be estimated
roughly at $m=m_0$. The following particular choice
$m_0=\sqrt{A^{-2}+B^{-2}+C^{-2}}$ significantly simplifies all
analytical expressions.

Taking into account that $m\gg m_0$, $\Pi$ is estimated to be
extremely small in the two-body-like case: $\Pi\ll(\omega
r)^2=v^2(r/R)^2\ll1$, where $r$ and $R$ are characteristic scales
(radial ($R\approx A$) and transverse ($r\approx B\approx C$)) of
the matter distribution projected on the rotation plane, and
$v~(=\omega R\ll1)$ is the orbital velocity.


The kinematic factor $v^2$ in $\Pi$ results in that the
quad\-rupole gravitational radiation occurs at the order of $v^7$
beyond Newtonian gravity instead of $v^5$ in the other approaches
with the partial violations of general covariance. The geometric
factor $(r/R)^2$ suppresses the energy radiation to zero for
point-like masses ($r\to0$) in accordance with the equations for
geodesics.

Tipical values of $\Pi$ are estimated to be $\Pi\ll10^{-2}$ for the
final stage of binary neutron stars ($v\sim0.3$, $r/R\sim0.5$),
and $\Pi\ll10^{-16}$ for the current stage of the binary pulsar
PSR1913+16. These estimations for $\Pi$ strongly contradict to
$\Pi=1$ in (\ref{EQF}), and the binary pulsar data.

A possible reason for the dramatic difference between the covariant
(\ref{GIQF}) and noncovariant (\ref{EQF}) estimations of the energy
loss rates is the covariance of the superenergy approach which
allows extracting only the covariant contributions from the
energy-momentum pseudotensor $t^{\mu\nu}$. This pseudotensor
contains contributions of inertial forces (with
$R_{\alpha\beta\mu\nu}=0$) which may dominate in $t^{\mu\nu}$ in
some particular gauges. Nevertheless, these inertial forces do not
contribute to the covariant energy flux of the real gravitational
fields with $R_{\alpha\beta\mu\nu}\ne0$. The strong contradiction
between the covariant and noncovariant calculations may mean
far-reaching consequences for the structure of any metric theory of
gravitation which claims to be realistic.\\

Acknowledgement: Author is grateful to S. I. Blinnikov, I. B.
Khriplovich, M. Yu. Konstantinov, and Yu. S. Vladimirov for their
constructive comments. This work was supported in part by research
grant \# 1885.2003.2 of Russian Ministry of Industry and Science.

\end{document}